\documentclass[journal]{vgtc}                     %

\newcommand{\add}[1]{\textcolor{black}{#1}}

\onlineid{1100}

\vgtccategory{Research}

\vgtcpapertype{Theoretical \& Empirical}

\title{
Causal Priors and Their Influence on \\ Judgements of Causality in Visualized Data
}

\author{
  \authororcid{Arran Zeyu Wang}{0000-0002-7491-7570}, \authororcid{David Borland}{0000-0002-0162-4080}, \authororcid{Tabitha C. Peck}{0000-0002-3667-7713}, \authororcid{Wenyuan Wang}{0000-0001-8765-6675}, and \authororcid{David Gotz}{0000-0002-6424-7374}
}

\authorfooter{
  \item Arran Zeyu Wang, Wenyuan Wang, and David Gotz are with the University of North Carolina at Chapel Hill (UNC). 
  E-mail: zeyuwang@cs.unc.edu, vaapad@live.unc.edu, gotz@unc.edu
  \item David Borland is with RENCI at UNC.
  E-mail: borland@renci.org
  \item Tabitha C. Peck is with Davidson College.
  E-mail: tapeck@davidson.edu
}

\shortauthortitle{Causal Priors}

\teaser{
\centering
  \vspace{-2mm}
\includegraphics[width=0.862\columnwidth]{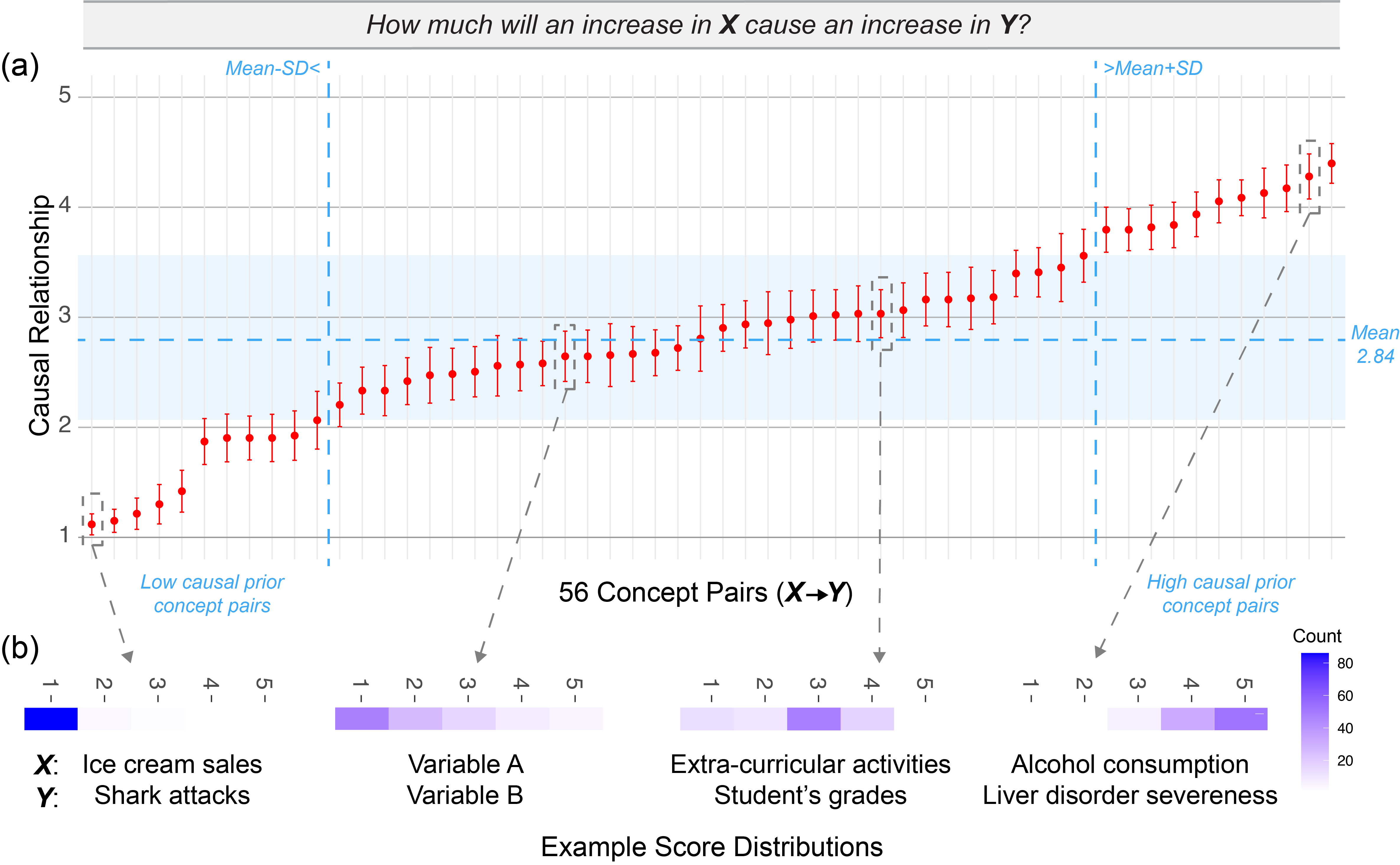}
  \vspace{-3mm}
\caption{ 
Results \add{from the first study in this paper} of participant-rated causal relationships for 56 concept pairs curated from open-source datasets.
Participants were asked questions in the form ``How much will an increase in X cause an increase in Y?'' for each X$\rightarrow$Y concept pair.
In (a), the Y-axis represents the participant-reported scores for concept causal relations (1 to 5: none to high). Each \textbf{concept pair} is placed in order by mean causal relation along the X-axis, showing 95\% confidence intervals.
The light blue horizontal band represents the mean score across all concept pairs $\pm$ one standard deviation (SD).
The \textbf{vertical dashed lines} delineate concept pairs that we refer to as having either low causal priors ($<$mean-SD) or high causal priors ($>$mean+SD).
Part (b) shows four example concept pairs from different parts of the causal prior spectrum. The heat maps show 
the number of participants in our study reporting each score on the 1-5 causal scale. As these results show, causal priors can vary widely across different concept pairs.
\add{In the second study in this paper (\autoref{sec-study2}), we examine the impact of these causal priors on visualization interpretation.}
  \vspace{-2mm}
}
\label{fig:teaser}
}
 
\abstract{
``\emph{Correlation does not imply causation}'' is a famous mantra in statistical and visual analysis.
However, consumers of visualizations often draw causal conclusions when only correlations between variables are shown. In this paper, we investigate factors that contribute to causal relationships users perceive in visualizations. We collected a corpus of concept pairs from variables in widely used datasets and created visualizations that depict varying correlative associations using three typical statistical chart types. We conducted two MTurk studies on (1) preconceived notions on causal relations without charts, and (2) perceived causal relations with charts, for each concept pair. Our results indicate that people make assumptions about causal relationships between pairs of concepts even without seeing any visualized data. Moreover, our results suggest that these assumptions constitute causal priors that, in combination with visualized association, impact how data visualizations are interpreted. The results also suggest that causal priors may lead to over- or under-estimation in perceived causal relations in different circumstances, and that those priors can also impact users' confidence in their causal assessments. In addition, our results align with prior work, indicating that chart type may also affect causal inference. Using data from the studies, we develop a model to capture the interaction between causal priors and visualized associations as they combine to impact a user's perceived causal relations. In addition to reporting the study results and analyses, we provide an open dataset of causal priors for 56 specific concept pairs that can serve as a potential benchmark for future studies. We also suggest remaining challenges and heuristic-based guidelines to help designers improve visualization design choices to better support visual causal inference.
  \vspace{-1mm}
} %

\keywords{Causal inference, Perception and cognition, Causal prior, Association, Causality, Visualization
  \vspace{-3mm}}

\graphicspath{{figs/}{figures/}{pictures/}{images/}{./}} %

\usepackage{tabu}                      %
\usepackage{booktabs}                  %
\usepackage{lipsum}                    %
\usepackage{mwe}                       %

\usepackage{mathptmx}                  %
\usepackage{amsmath,amsfonts}
\usepackage{algorithmic}
\usepackage{algorithm}
\usepackage{array}
\usepackage{textcomp}
\usepackage{stfloats}
\usepackage{url}
\usepackage{verbatim}
\usepackage{graphicx}
\usepackage{cite}  
\usepackage{amsmath}

\begin{document}

\maketitle
\section{Introduction}
\label{sec-intro}

Supporting causal inference for users is a foundational pursuit for visualization and visual analytics~\cite{borland2024using, tukey1977exploratory, kim2020bayesian, hullman2021designing, zhou2021modeling}. However, it is a challenging task and the assumption of causal relationships where there may be none can lead to significant misinterpretations, affecting decision-making processes across various domains~\cite{pearl2009causality}.

Well-known analytical guidelines caution against conflating correlation with causation. However, when viewing visualizations (which typically display correlative associations rather than causal relations), users often make assumptions regarding causality.
Effectively communicating causal relationships, or the lack thereof, remains challenging for data visualizations~\cite{borland2024using}.
Understanding how humans interpret causal relationships from visualized data is therefore crucial for effective visualization design.

Acknowledging that understanding relationships---causal or otherwise---is essential, this paper aims to gain insights into how humans infer causal relationships when they are consuming visualizations, a process that we refer to as \emph{visual causal inference}.
Previous studies have scrutinized how users examine the data presented in a visualization as well as the manner of visual presentation to understand how these factors influence users’ visual causal inferences~\cite{wang2024empirical, kale2021causal, xiong2019illusion}. 
Concurrently, research into human perception has shed light on how it shapes the interpretation of visual information of associations and correlations~\cite{cleveland1988shape, rensink2010perception}. 
However, humans' visual causal inferences do not occur in isolation; they are made within a broader context that includes specific choices of tasks and may be influenced by cognitive effects.

Beyond these existing contextual factors, we theorized that underlying semantic causal priors---a person's preconceived notions about concepts and the underlying causal relationships between them~\cite{rudolph1997psychological}---play a critical role in the causal inferences drawn from charts.
We therefore set out to investigate whether and how these underlying causal priors might affect the causal inferences people make (i.e., whether the same chart can lead to different interpretations under different causal priors).

This paper introduces two related studies aimed at answering the questions of (1) whether underlying causal priors influence causal inferences from visualizations, and if yes, (2) how these priors interact with the actual statistical associations depicted in visualizations presented with different chart types.

We first collected a corpus of concept pairs by selecting variables from widely-used datasets and created visualizations representing varying statistical association levels with three typical statistical chart types: scatterplots, line charts, and bar charts, \add{representing continuous, temporal, and categorical data respectively}.
We then conducted two crowdsourced studies to: (1) measure the underlying causal priors for the concept pairs in our corpus as rated by participants when not seeing any visualization of data at all; and (2) examine the effects of the causal prior, visualized association, and chart type on participants’ perceptions of causal relations in visualized data.
Our results help provide a deeper understanding of how preconceived notions about causality combine with visual evidence to influence human cognition of causal inference. We analyze these results in light of previous studies that looked at human perception of causal inference in visualizations, and suggest actionable heuristic-based design guidelines for visual causal inference.
\add{Finally, we highlight some remaining challenges 
such as the need for a deeper understanding of the impact of specific chart and data types.}

More specifically, the contributions of this paper include:
\begin{itemize}
    \item {\bf Empirical evidence and reference measurements of causal priors associated with  concept pairs.} We report data from a study showing that people infer causal relationships between pairs of concepts absent of any other information or charts. Data from this study is provided as an open dataset of priors for 56 concept pairs that can serve as a benchmark for future studies.

    \item {\bf Empirical evidence demonstrating the effect of causal priors on users' visual causal inferences.} We report data from a study showing that a user's assessment of causality when viewing a chart is influenced not only by the type of chart and the data being visualized, but also by the causal prior.
    
    \item {\bf A model capturing the interaction between causal priors and other factors.} Based on empirical data gathered in our studies, we define a model that expresses the interaction between causal priors, visualized association, and chart type in influencing the user's reported strength of the causal relationship between visualized concepts.

    \item {\bf A discussion of implications and heuristic-based design guidelines.} A detailed analysis of the study results is provided along with a discussion of the implications of these results on our understanding of visual causal inference. This includes a set of heuristic-based guidelines to help designers improve visual design choices to better support visual causal inference.

\end{itemize}

\section{Background and Related Work}
\label{sec-related}

The research presented in this paper builds upon prior work in two broad areas of research: \add{i) human inferences about causal relationships based on viewing visualizations of data and ii) the role of a person's prior context or knowledge in human cognition.}

\subsection{Human Perception in Visual Causal Inference}

The term \emph{causal inference} is typically used to refer to techniques for identifying and characterizing relationships between variables that are indicative of cause and effect~\cite{pearl2009causality}.
The mathematical foundations of these approaches have been developed within the 
statistics and machine learning communities, where the techniques have been widely applied~\cite{pearl2009causality, spirtes2000causation, prosperi2020causal}. 

Motivated in part by these advances, causal inference has also drawn attention as an emerging topic in visualization research~\cite{borland2024using, hullman2021designing, kim2020bayesian}. For example, a number of visualization techniques have been proposed to support specific causal inference workflows, such as event sequence or time-series analysis~\cite{wang2022domino, jin2020visual, deng2021compass}, paradox and illusion detection~\cite{wang2015visual, salim2024belief, wang2023countering}, algorithm interpretation~\cite{hoque2021outcome, ghai2022d}, and general exploratory tasks~\cite{yen2019exploratory, wang2024framework, kaul2021improving, guo2023causalvis}. See Borland et al. for a more detailed discussion on recent advances in developing visualization techniques that can directly support visual causal inference~\cite{borland2024using}.

A number of these techniques are designed to provide graphical representations and controls over an underlying \emph{statistical causal inference} process. Others, however, examine or support the ways in which humans cognitively draw inferences about causal relationships when looking at data visualizations. We emphasize that this idea of a user's \emph{visual causal inference} is closely related to but distinct from statistical inference approaches. Visual causal inferences are ones made by a user based in part on their perception and interpretation of the data they see during an analytical task.

Critically, previous research has shown that users make visual causal inferences even when viewing basic statistical charts that may focus on correlation or other non-causal statistical measures (e.g.,~\cite{kaul2021improving}). For this reason, the understanding of how human perception relates to the ways that people draw visual causal inferences has become an active topic of research~\cite{borland2024using}.

For example, Xiong et al. found that different visualization types do not equally support causal interpretation~\cite{xiong2019illusion}. For instance, their results suggest that bar charts may be more suggestive of causal relationships than scatter plots.
The same research also explored the impact of specific visual encodings (as variants of the same basic chart type) and found that increasing levels of data aggregation were associated with increasing levels of perceived causality. Similarly, the same study suggests that within a given type of chart, lines and dots were viewed as more suggestive of causality than bars.

In other work, Kale et al.~\cite{kale2021causal} employed mathematical psychology~\cite{griffiths2005structure} and a causal support model to study causal inference perception with visualizations.
Their model design supports comparisons between that ground truth and users' perceptions compared to several other works~\cite{xiong2019illusion, kaul2021improving}.
They found that users' causal inferences deviated from the ground truth, with either overestimation or underestimation of the underlying ground truth causal relationships.
The results of this study highlight the difficulty in precisely assessing the level of evidential support that a particular dataset offers for a specific causal explanation.

Kaul et al. found that users tend to infer causal relations when filtering data using variable constraints, employing simple bar charts and line charts to show correlation~\cite{kaul2021improving}.
Moreover, they found that using counterfactual visualizations to explore different data subsets can partially mitigate these effects when causal relationships are unsupported by the data.
Building upon these findings, Wang et al. developed a causality comprehension model for counterfactual-based visualizations~\cite{wang2024empirical}. They also reported results from a study that showed that visualizing counterfactuals within static charts provides benefits to users at various levels of causal comprehension.

Taken together, these studies provide significant insights into how people infer causal relationships based on their perceptions of various elements of how the data is visually represented. This focus is well-motivated and the effects of human perception are important to understand. However, in practice, perception only plays one part in the broader cognitive process of drawing visual causal inferences. 
In this paper, we aim to begin addressing this gap by exploring how certain cognitive factors---specifically those related to assumptions made by users based on semantics associated with relationships between concepts---impact how people make visual causal inferences.

\subsection{Human Cognition and Underlying Causal Priors}

Human knowledge and assumptions about the world can be complex and varied, informed by each person's lived experience and in response to various forces such as culture and training~\cite{koffka2013principles, szafir2023visualization}. This is particularly true in how people understand language, as reflected in the results from several studies that have explored how people assume relationships for and between concepts in human communication. For example, research has long shown that different English verbs communicate different levels of implied causality in short phrases, a result that has been replicated in several studies~\cite{rudolph1997psychological}. This same phenomenon was more recently replicated by Ferstl~et al. in research that also produced an annotated corpus of 305 English verbs based on data gathered during the study~\cite{ferstl2011implicit}. 

Demonstrating that implied causality is not unique to English,  Goikoetxea et al. studied the implied causality communicated by 100 interpersonal verbs in Spanish~\cite{goikoetxea2008normative}. Like the previously mentioned studies of English verbs, this study found that different verbs implied different types and degrees of causal relationships. 
 Moreover, they found that the results were similar for both adult and child participants.

These findings from the psychology and human behavior literature have led to related attempts within the NLP community to identify implied causal relationships within text corpora. For example, Riaz and Girju proposed an NLP model to recognize causality in verb-noun pairs by predicting words' semantics~\cite{riaz2014recognizing}. In another example, Salim et al. developed the \emph{BeliefMiner} system to extract causal graph networks from unstructured text and used the approach to explore causal illusions about climate change~\cite{salim2024belief}.

These efforts demonstrate that people tend to implicitly associate causal relationships with short phrases or individual words. These implied causal properties exist even before those concepts might appear in a visualization, for example as part of an axis, legend, or some other chart annotation. We refer to this idea---that a causal relation can be implied by the use of words in and of themselves, even before those words are used within a visualization---as \emph{causal prior}.

In this way, causal priors are similar to other factors describing prior experience or knowledge that have been shown within the visualization literature to influence how people interpret charts~\cite{koffka2013principles, szafir2023visualization}.
For example, research has shown that both background knowledge and individual differences can influence visualization comprehension~\cite{quadri2024do}.
The efficiency of specific visualization designs depends in part on characteristics of the potential audience such as internal representations~\cite{larkin1987diagram}, graphical literacy~\cite{franconeri2021science}, color understanding~\cite{schloss2018mapping, szafir2018modeling}, mental models~\cite{liu2010mental}, diverse backgrounds~\cite{peck2019data, hall2021professional, wang2024automated}, experience levels~\cite{grammel2010information, burns2023we}, and cross-domain knowledge gaps~\cite{padilla2018decision}.
Most recently, Xiong et al. reported results that show that users' personal beliefs can bias their estimates of visualized correlations~\cite{xiong2022seeing}. 

Yet despite this breadth of research exploring how users' priors influence their interpretations of visualizations, to our knowledge, there have been no previous studies that have examined the effect of causal priors on how people perform visual causal inference using visualizations.
In this paper, we both: (1) develop a corpus of concept pairs annotated with causal priors; and (2) study how those priors influence users' visual causal inference behavior in different visualization contexts.

\section{Methodology}
\label{sec-method}

To investigate the existence and influence of causal priors, we designed and conducted two distinct yet complementary empirical studies by recruiting users from Amazon’s Mechanical Turk to assess human cognition of causal inference. Both studies were approved by the [Redacted] Institutional Review Board.
The studies are designed to investigate potential visual and non-visual impact factors on human cognition when conducting causal inference

\subsection{Terminology}
\label{sec:terminology}

Here we provide definitions for three key terms used throughout the paper:

\begin{itemize}
    \item \textbf{Causal prior:}
    The causal relation level of a directed pair of concepts, as indicated by participants in Study 1  without any visualization stimulus.
    Values exist on a scale from 1 (no relation) to 5 (high causal relation).
    
    \item \textbf{Visualized association:}
    The statistical association level between a pair of concepts that is represented in a visualization. This association level is controlled by the visualization stimulus generation process used for Study 2 (see \autoref{sec:visstimuli}), comprising five levels from 0 to 1 with intervals of 0.25.
    
    \item \textbf{Perceived causal relationship:}
    The degree of causal relationship between concept pairs perceived by a user when viewing a visualization.
    In this paper, the values refer to our results from Study~2. As with Study 1, values exist on a scale from 1 (no relation) to 5 (high causal relation).
    
\end{itemize}

\subsection{Hypotheses}

In this paper we aim to explore the potential relations among causal priors, visualized associations, and perceived causal relationships.
Based on this goal, we hypothesized that:

\begin{itemize}
    \item \textbf{H1:} People have underlying causal priors reagarding the assumed strength of causal relationships for concept pairs.
    \item \textbf{H2:} Causal priors affect the perception of causal relationships from charts.
    \item \textbf{H3:} Visualized associations affect perceived causal relationships from charts.
    \item \textbf{H4:} The nature of the effects of H2 and H3 can, in part, be explained by the \textit{disagreement} between causal priors and visualized associations.
    \item \textbf{H5:} The impact of causal priors and visualized associations on perceived causal relationships vary by chart type.
\end{itemize}

Through these hypotheses, our study aims to contribute to the broader discourse on graphical perception, specifically in the context of how human cognition interprets causality from visual data. 
The exploration of these hypotheses will not only enhance our understanding of cognitive processes, but also inform the design of more effective visualization-guided exploration systems for causal inference.

\subsection{Study 1: Examining Causal Priors for Concept Pairs}
Study 1 examines inherent beliefs about causality by measuring the prior causal relationship ascribed to a given directed concept pair.
Participants in this study were asked to assess the strength of causal relationships between concept pairs on a 5-point Likert scale based solely on the concept names, without any visualizations. 
Instances of the questions are provided in \autoref{sec:causalquiz}. Participants also provided their confidence in their causal strength ratings on a 5-point Likert scale.
Each participant saw each concept pair in a randomized order.
Study 1 was designed to address \textbf{H1}, and provide the causal priors used to assess \textbf{H2-H5} in Study 2.

\subsubsection{Participants}

A cohort of 100 participants was recruited for Study 1, with an average engagement time of 12 minutes, leveraging the Amazon MTurk platform using CloudResearch with at least a 95\% approval rating and IP addresses from the United States and Canada.
These selection criteria aimed to ensure a certain level of homogeneity in cultural and educational background, which can influence cognitive processing.
We excluded eight participants who failed at least two random attention checks (these excluded participants spent less than 3 minutes on average), and analyzed data from the remaining 92 participants, resulting in a 92\% acceptance rate.
The final study 1 participants included 65 men and 27 women, ranging from 24--58 years of age.

\subsubsection{Dataset and Stimuli}
\label{sec:stimuli}

Our empirical investigation employed a self-collected corpus comprising two related components: concept pairs, and causal questions.

\paragraph{Concept Pairs:}
\label{sec:variablepairs}
A total of 56 concept pairs were curated for our study.
We first collected 50 pairs of variable names from ten widely used open-source
datasets, including the \emph{Titanic},
\emph{Census Income}, and \emph{Heart Disease} datasets.
These datasets were selected from the ``Popular Datasets'' list on the 
UCI data repository~\cite{asuncion2007uci} and the ``Trending Datasets'' list on Kaggle in an attempt to cover a diverse range of concepts.
Concept names for selected variables were determined by the authors, and then reviewed by five individuals with a maximum education level of a high-school diploma or equivalent to confirm that the names could be interpreted by the general public.
We slightly revised the concept names from the original datasets based on their feedback, e.g., ``computer speed'' was used to simply replace ``RAM size'' which can be more complex to understand.
These 50 pairs were supplemented with an additional 5 concept pairs, such as
\textit{divorce rate in Maine} and \textit{per capita consumption of margarine},
that were selected from Spurious Correlations~\cite{vigen2015spurious} to include pairs that we expected to be rated as having little or no causal relationship.
Finally, we also included \textit{Variable A} and \textit{Variable B} to establish a baseline for a non-semantic concept pair.

For each concept pair, we separated them into one causal factor and one outcome: we employed the widely used factor-outcome relations of the 50 pairs from datasets (e.g., studying time can be a causal factor to impact students' grades), followed the original order of the 5 pairs from spurious correlations~\cite{vigen2015spurious}, and determined \textit{Variable A} as the causal factor for the non-semantic concept pair.
The causal factor is denoted as \textbf{X} (e.g., \textit{Alcohol consumption} in \autoref{fig:teaser} (b)), and the outcome as \textbf{Y} (e.g., \textit{Liver disorder severeness} in \autoref{fig:teaser} (b)). This assigned a causal direction to each concept pair, denoted \textbf{X}$\rightarrow$\textbf{Y}, and participants were only asked to consider the causal relationship for each concept pair in the specified direction.

\paragraph{Study Questions:}
\label{sec:causalquiz}
Each concept pair was accompanied by a question
designed to determine each participant's rating of the strength of the causal relationship between the two concepts, e.g., ``How much will an increase in \textbf{computer speed} cause an increase in \textbf{laptop price}?''.
All questions were asked with an assumption of a positive causal relationship, i.e., how much does an increase in one cause an increase in the other.
The questions were designed to provide sufficient background context to answer the questions, including phrases such as ``in the Titanic sinking'' or ``in a cancer surgery.''
Further, we validated the questions with five individuals who hold high-school level or lower educational degrees to ensure the questions would be understandable for general users.
Participants were asked to rate the strength on a 5-point Likert scale from 1 (no causal relationship) to 5 (high causal relationship). Participants were also asked to assess their confidence in each rating on a 5-point Likert scale from 1 (low confidence) to 5 (high confidence).
In addition, three simple attention check questions were added, e.g.,~``Please choose the answer of 3+7.'', displayed in random order and position within the first 1/3, second 1/3, and last 1/3 of all questions.

Details of concept pairs, used datasets, and study questions are all available in the supplemtns on \href{https://osf.io/dfkv4/?view_only=f84ffbc28cdf45e5a3d68f2f1e9c8427}{\textcolor[RGB]{0,0,255}{OSF}}.

\subsection{Study 2: Exploring the Effects of Visualizations on Causality Perception}
\label{sec-study2}

\begin{figure}[t]
    \centering
    \includegraphics[width=\columnwidth]{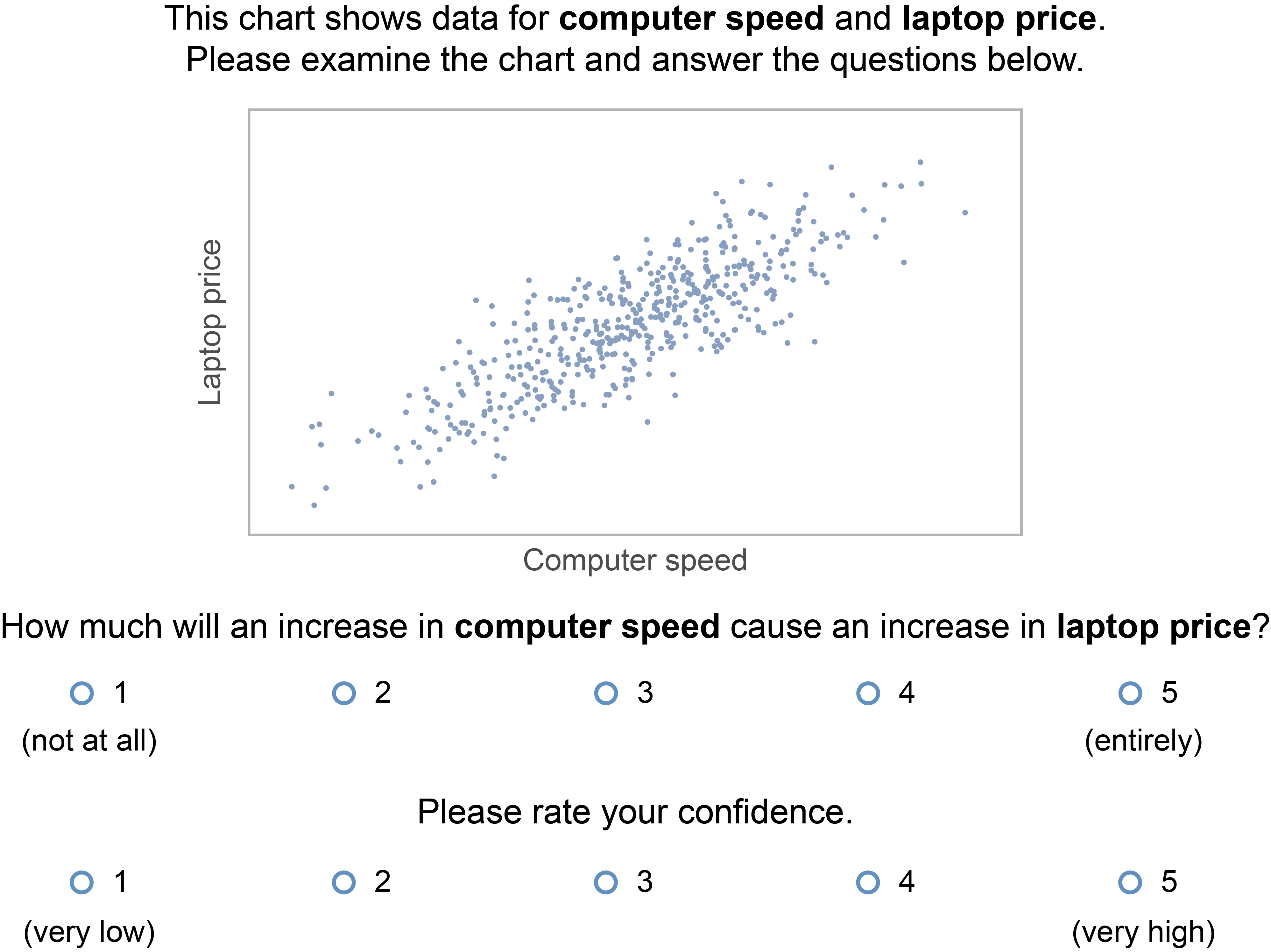}
    \caption{An instance of a visualization stimulus combined with a causal question and a confidence question from Study 2.
    }
    \label{fig:chart}
    \vspace{-1em}
\end{figure}

Study 2 aimed to investigate how the perception of causal inference is influenced by the causal priors from Study 1 as well as varying levels of visualized associations in charts.
The tasks in Study 2 mirrored those of Study 1, however a visualization for the concept pairs was also provided.
Participants reported perceived causal relationships based on these visualizations, which
varied by three chart types (line, bar, scatterplot) and different levels of visualized correlative association (ranging from 0 to 1).
Duplicate charts or questions were avoided, and neither the chart type nor the association level was repeated consecutively during the study to avoid potential biases and learning effects.
A representative chart example is provided in \autoref{fig:chart}.
Study 2 was designed to address \textbf{H2-H5}. 

\subsubsection{Participants}

For Study 2, we recruited a new cohort of 250 participants again using Mechanical Turk and the same user restrictions as Study 1. Participants who completed Study 1 were ineligible to complete Study 2. 21 participants were excluded from the final analysis due to failing at least 2 out of 3 of the attention checks, resulting in a 91.6\% acceptance rate. The final Study 2 participants included 138 men, 87 women, and 4 non-binary, ranging from 22--64 years of age. The experiment took approximately 15 minutes.

\subsubsection{Visualization Stimuli}
\label{sec:visstimuli}
In Study 2 we employed the same set of concept pairs and study questions as Study 1.
We classified the concept pairs into three categories based on their inherent data types: time series, categorical, and continuous.
\add{To balance study design simplicity with a reasonably representative sample of commonly-used visualizations, we chose three of the most widely applied chart types~\cite{quadri2021survey} for the three data types: line charts for time series, bar charts for categorical, and scatterplots for continuous.
These three chart types have also been commonly included in various types of graphical perception studies such as visualization comprehension~\cite{quadri2024do}, counterfactual visualization~\cite{wang2024empirical}, color perception~\cite{szafir2018modeling}, and causality visualization~\cite{xiong2019illusion}.}
The breakdown of concept pairs per chart type is as follows. The line chart included 16 concept pairs from open-sourced datasets, the non-semantic concept pair, and the five spurious concept pairs~\cite{vigen2015spurious} for a total of 22 line chart trials. The bar chart and scatterplot each included 17 concept pairs from open-sourced datasets, for a total of 17 trials each. 

\begin{figure}[h]
    \centering
    \includegraphics[width=\columnwidth]{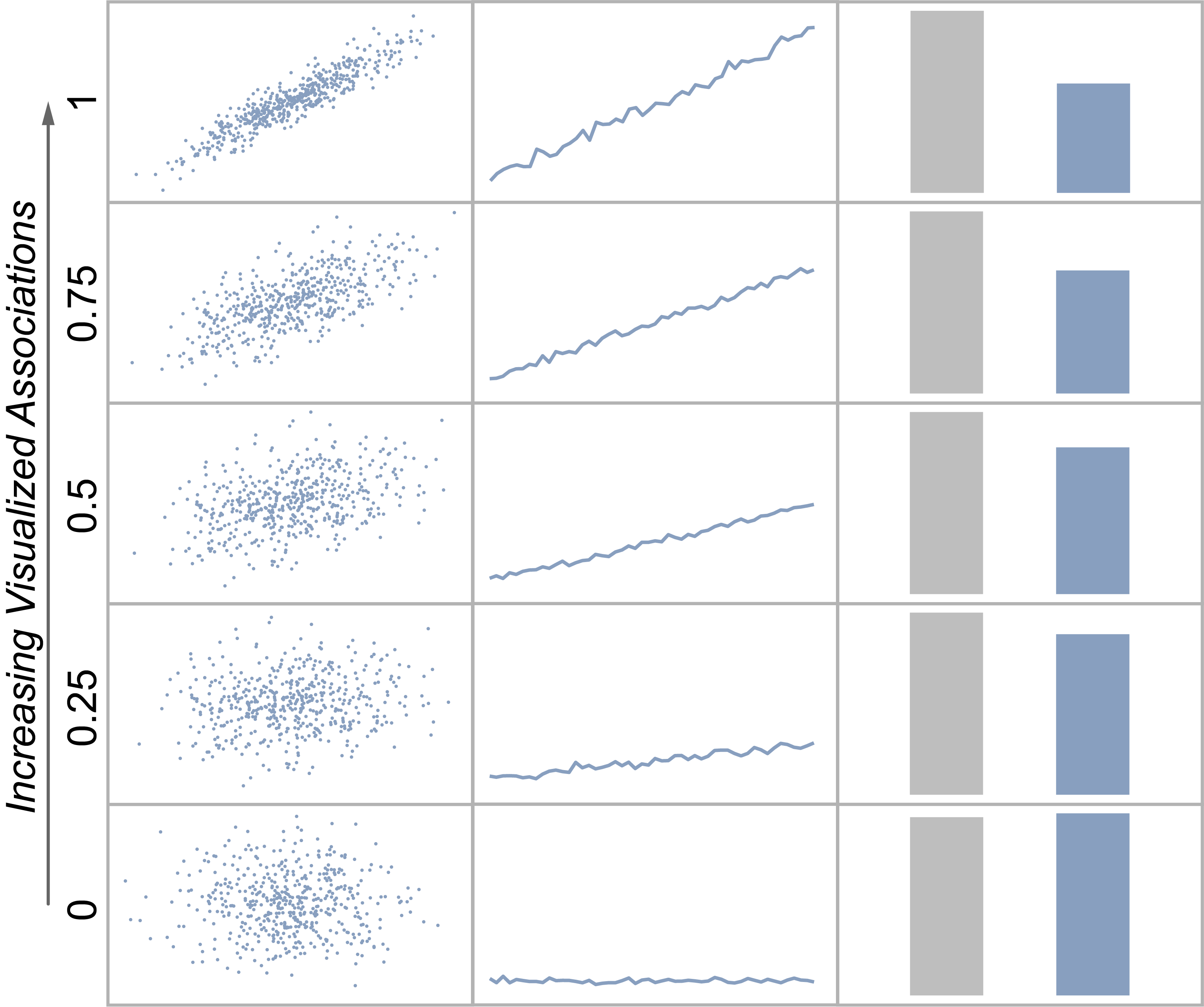}
    \caption{\add{This figure shows illustrations of the three chart types and five visualized association levels employed in Study 2.} Left to right: scatterplots, line charts, and bar charts. Bottom to top: low association (around 0) to high association (around 1).}
    \label{fig:level}
    \vspace{-1em}
\end{figure}

To explore the impact of causal priors on perceived causal relationships, we displayed data in each chart type using five distinct association levels (\autoref{fig:level}):

\noindent \textbf{Scatterplots:} We adjusted the multivariate covariance, with values spanning from 0 (weak association) to 1.0 (strong association), at intervals of 0.25.
Following existing research on correlation judgments in scatterplots~\cite{rensink2010perception, rensink2017nature, cleveland1986experiment}, we only tested positive correlations to simplify the study's complexity and avoid the impact of negative correlations.

\noindent \textbf{Line charts:} Previous studies on line charts~\cite{cleveland1988shape, heer2006multi, wang2017there, wang2018image} show that the effectiveness of human perception is highest at around a 45\textdegree{} trend.
We therefore generated 5 evenly spaced association levels from 0 to 1 with a linear mapping of \add{\emph{slope values} (i.e., the average y to x ratio of a line chart)} from 0\textdegree{}  (no association) to 45\textdegree{}  (strong association).

\noindent \textbf{Bar charts:} We chose simple two-bar charts following the methodology from prior work~\cite{xiong2019illusion} by manipulating the difference ratio between the bar heights to control the association difference. Differences for the 5 association levels ranged from no difference at 100\% to 100\%, to a maximum difference of 100\% to 50\%, with evenly spaced intervals in between. 

Additionally, the visualized associations \add{across all three chart types} were generated by adding random noise with a variance in the open range (0, 0.1) to avoid duplicated charts and unrealistic cases, e.g. a scatterplot with a covariance of 1.0 that would form a straight line.
\add{The upper threshold of random noise (0.1) was derived from previous work, in which differences of around 10\% were often chosen as a threshold for correlation or slope value comparisons for simple statistical charts~\cite{talbot2012empirical, rensink2017nature}.}
In this way, the difference between two neighboring association levels was close to 25\%.
In the following analysis, we refer to these charts using their base association values of 0, 0.25, 0.5, 0.75, and 1.

\add{These values are shared across all chart types. However, we acknowledge that the chart types are each associated with specific data types. This results in different distributions across chart types and makes interpretations of the impact of chart type more difficult. Please see \autoref{sec:analysis_chart} and \autoref{sec:discusschart} for more discussions.}

\subsection{Procedure}

Both Study 1 and 2 shared a similar procedure. The studies were administered using Qualtrics. Before beginning the study, participants completed an informed consent form, acknowledging their understanding of the study's purpose and their rights as participants.
Participants completed 56 randomized trials, one for each concept pair, and three attention check questions, for a total of 59 trials. For each concept pair, participants were asked, ``How much will an increase in \textbf{X} cause an increase in \textbf{Y}?" where \textbf{X} and \textbf{Y} were the concept pair. Participants answered on a 5-point Likert scale from 1 (not at all) to 5 (entirely). Participants were then asked, ``Please rate your confidence" on a 5-point Likert scale from 1 (very low) to 5 (very high). 

For Study 2, each concept pair question also included a visualization (\autoref{fig:chart}). To avoid learning and fatigue effects, each participant saw each concept pair once, with its associated chart type (see \autoref{sec:visstimuli}) showing a randomly chosen visualized association level. Visual association levels were evenly distributed for each participant such that each participant saw each association level 10-11 times. Consecutive trials prohibited repeated chart types and visual associations. E.g., a bar chart was never shown directly after another bar chat, and a visual association with level 1 was never shown directly after a visual association with level 1. 

Across all participants, each association level for each concept pair visualization was shown a total of 50 times. After excluding participants who failed attention checks, the total number of completed trials for each concept pair and association level varied from 44 to 48 trials each.
After completing the 59 concept pair and attention check trials, participants were asked to provide any feedback.
In the end, we noticed users the data shown in our study were synthetic and did not depict any real-world relationships.
On average, Study 1 took 12 minutes and Study 2 took 15 minutes.

\subsection{Analysis Overview}

In \autoref{sec-analysis-1} and \autoref{sec-analysis-2} we discuss significant results and statistical analysis for the two studies based on the independent factors considered in this paper with 95\% bootstrapped confidence intervals ($\pm$ 95\% CI) for fair statistical communication~\cite{dragicevic2016fair}.
The study data, analysis code, and the open dataset of priors are available at \href{https://osf.io/dfkv4/?view_only=f84ffbc28cdf45e5a3d68f2f1e9c8427}{\textcolor[RGB]{0,0,255}{OSF}}.

\section{Study 1 Analysis}
\label{sec-analysis-1}

To test \textbf{H1} we performed a repeated measures ANOVA on the 56 concept pairs.
Our analysis reveals a significant effect of different concept pairs on causal rating ($F(55, 5152)=53.41, p<.0001, \eta^2=.36$), indicating that people assume different causal relationship strengths for different concept pairs.

We then calculated the average value of users' reported causal relationships for each concept pair to serve as \add{a key factor,} causal priors, used in the analysis of Study 2.
\autoref{fig:teaser} shows these average values with 95\% confidence intervals for each concept pair.

The mean causal relationship strengths indicate an obvious difference in the score distribution across different concept pairs (see \autoref{fig:teaser} (a)).
In addition, the relatively tight 95\% confidence intervals indicate good agreement across users for the causal relationship ratings.
The 5 spurious correlations also had the five lowest causal priors (e.g., see \emph{ice cream sales} and \emph{shark attacks} in \autoref{fig:teaser}), further justifying the design.  The results from Study 1 therefore support \textbf{H1}, indicating that people have underlying causal priors between specific concept pairs.

\section{Study 2 Analysis}
\label{sec-analysis-2}

Given the experimental support for underlying causal priors shown in Study 1, we include the average causal relationship strength per concept pair from Study 1 as a variable, \textit{causal prior}, to the analysis in Study 2.

\add{A generalized multiple linear regression model} was used to analyze the \textit{perceived causal relationship} based on \textit{causal prior}, \textit{visualized association}, and \textit{chart type}. \add{This approach can be better balanced for the distribution differences between chart types as mentioned in \autoref{sec:visstimuli}}. The regression of \textit{causal prior}, \textit{visualized association}, and \textit{chart type} on \textit{causal relationship} was statistically significant, ($F(11,14404)=257$, $p<.0001$, $R^2=.17$).
\autoref{tab:anovas2} summarizes the significance results from the linear regressions for \textit{perceived causal relationship}, which are presented in more detail in the following sections.

\begin{table}[htbp]
\centering
\caption{\add{Linear regression results from Study 2 for \textit{perceived causal relationship}.} Significant effects are indicated in bold and the corresponding rows are highlighted in green. 
\add{Note that line charts are the reference of the other two chart types in the model so it cannot be shown below.}
}
\includegraphics[width=0.9\linewidth]{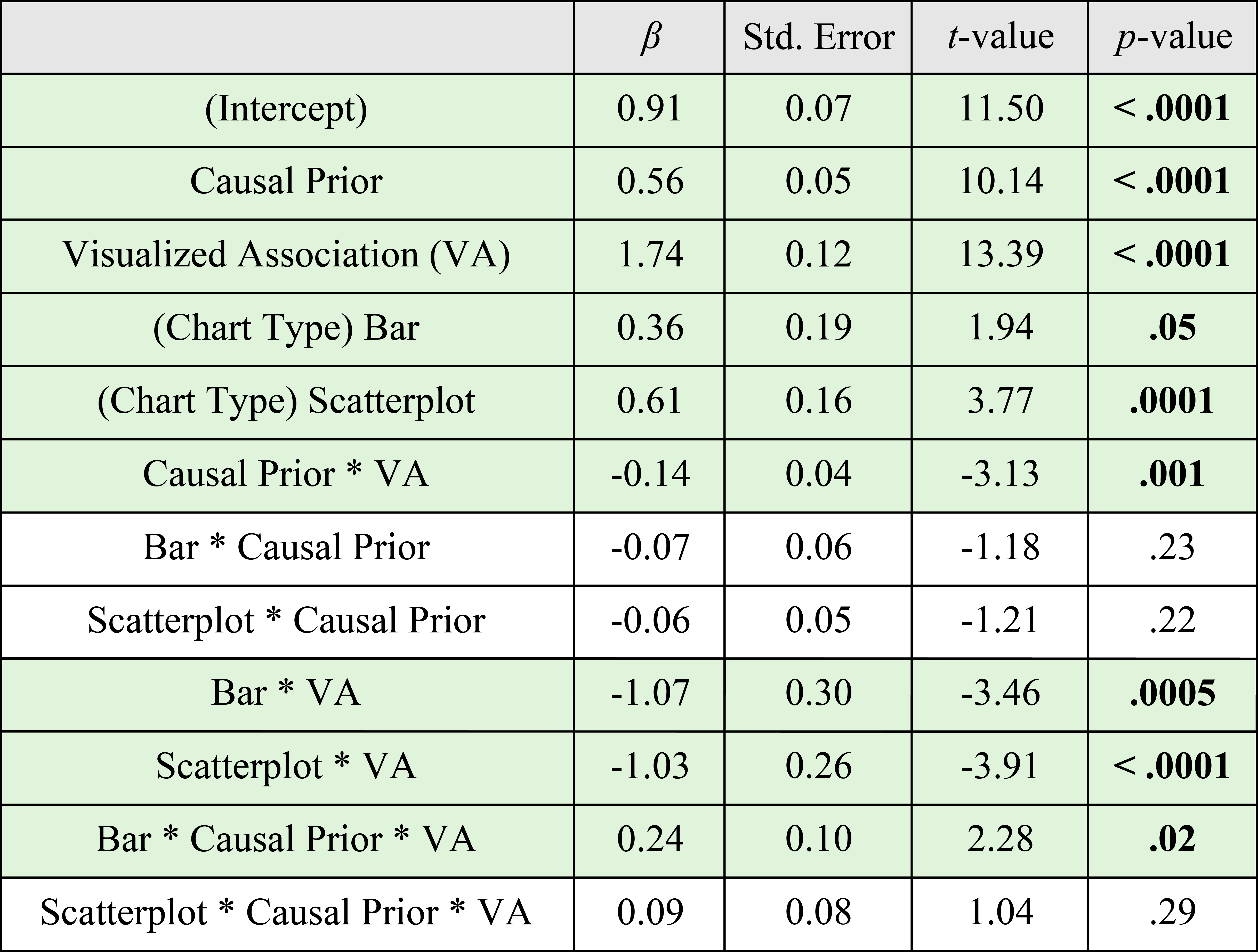} 
\label{tab:anovas2}
\vspace{-1em}
\end{table}

\subsection{The Impact of Causal Priors on Perceived Causal Relationships}
\label{sec:analysis_prior}

Our results support \textbf{H2}: Causal priors affect the perception of causal relationships
from charts. 
\textit{Causal prior} ($\beta=0.56$, $p<.0001$) significantly affects \textit{perceived causal relationship}. This supports that higher \textit{causal priors} were associated with higher ratings for the strengths of \textit{perceived causal relationships}.

\subsection{The Impact of Visualized Associations on Perceived Causal Relationships}
\label{sec:analysis_va}

Our results support \textbf{H3}: Visualized association affects perceived causal relationships from charts.
\textit{Visualized association} ($\beta=1.74$, $p<.0001$) significantly affects \textit{perceived causal relationship}. The result indicates that higher \textit{visualized association} is associated with higher ratings for the strengths of \textit{perceived causal relationship}.

\subsection{Interaction between Causal Priors and Visualized Associations}
\label{sec:analysis_agree}

Our results support \textbf{H4}: The nature of the effects of \textbf{H2} and \textbf{H3} can, in part, be explained by the disagreement between causal priors and visualized
associations.

\begin{figure}[b]
    \centering
    \includegraphics[width=0.75\columnwidth]{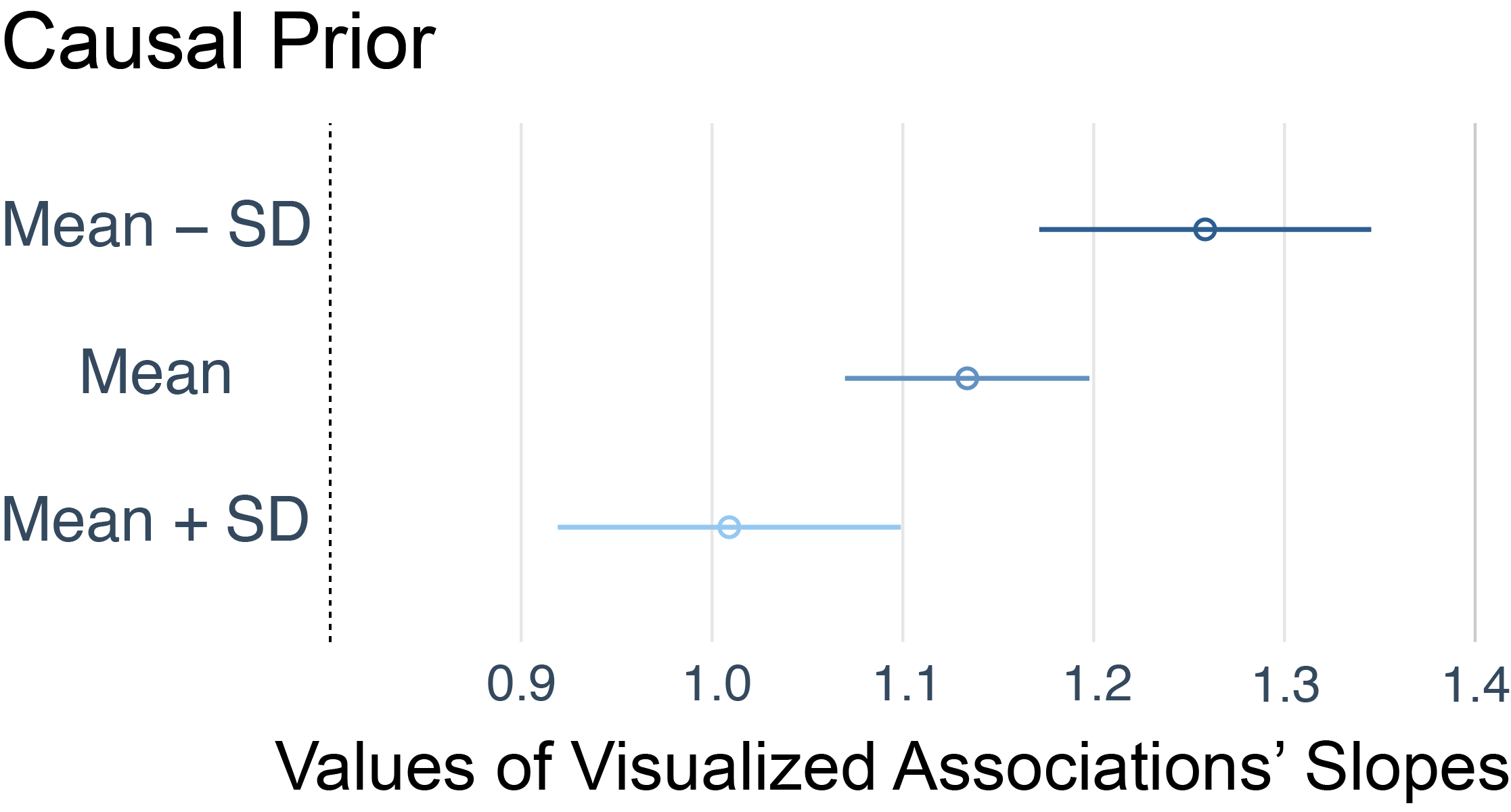}
    \caption{
    \add{Results of the simple slope analysis between causal priors and visualized associations.
    It shows that visualized associations would have a lower impact for higher causal priors.
    }
    From top to bottom, the dark blue line shows the result for the mean-SD causal prior (2.01), blue shows the result for the mean causal prior (2.84), and light blue shows the result for the mean+SD causal prior (3.66).
    }
    \label{fig:slope_va_p}
\end{figure}

A significant interaction was found between \textit{causal priors} and \textit{visualized associations} ($\beta = -0.14$, $p=.001$).
Post-hoc analysis of simple slopes was performed at the \textit{causal prior} mean $\pm$ standard deviation, in which $mean-SD = 2.01$, $mean = 2.84$, and $mean+SD=3.66$.
\add{Note that the slope differences here are distinct from the \emph{slope values} that were used to define associations in our line chart stimuli (\autoref{sec:visstimuli}). In this instance the slope differences from the simple slope analysis results indicate how the influence of \textit{visualized association} changes as the \textit{causal prior} changes.}

Significant differences were found between all slope pairs. There was a significant difference between \textit{mean-SD} (M=1.26, SE=0.04) and the \textit{mean} (M=1.14, SE=0.03), ($t(14052)=4.01$, $p=.0002$, $r=.82$), a significant difference between \textit{mean} and \textit{mean+SD} (M=1.01, SE=0.05), ($t(14052)=4.01$, $p=.0002$, $r=.84$), and a significant difference between \textit{mean+SD} and \textit{mean-SD}, ($t(14052)=4.01$, $p=.0002$, $r=.94$). %
See 
\autoref{fig:slope_va_p} for the visualization of simple slopes.

The results show that the \textit{mean-SD} has a significantly steeper slope than both the \textit{mean}  and the \textit{mean+SD}. This difference indicates that the \textit{visualized association} has a greater influence on the \textit{perceived causal relationship} at lower \textit{causal prior}. As \textit{causal prior} increases, the influence of \textit{visualized association} decreases. Though, for all cases, the slope is positive indicating that \textit{visualized association} positively impacts \textit{perceived causal relationship}.

\subsection{The Impact of Chart Type}
\label{sec:analysis_chart}

\begin{figure}[t]
    \centering
    \includegraphics[width=\columnwidth]{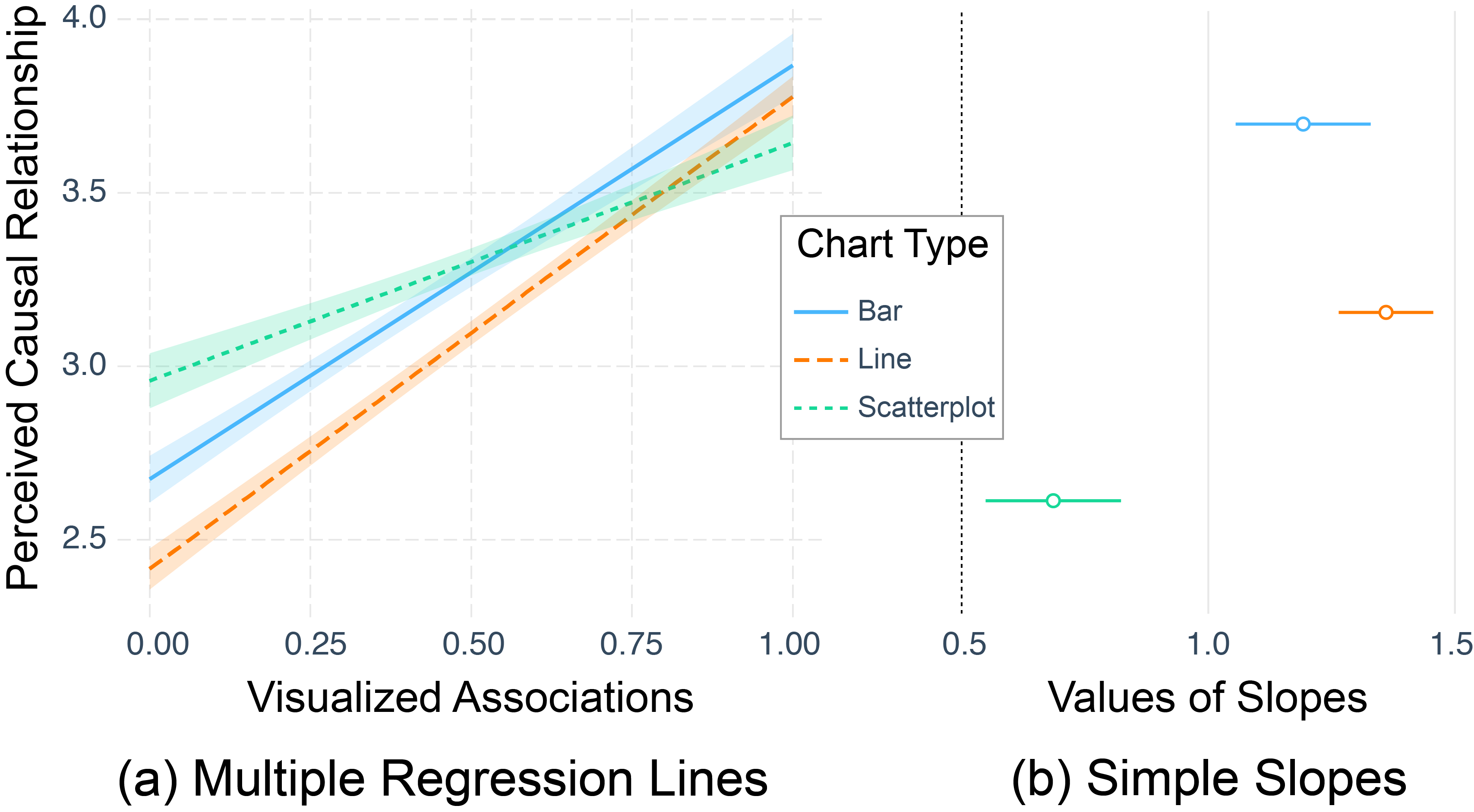}
    \caption{
    \add{Results of the simple slope analysis between chart types and visualized associations.
    The values hint scatterplots have a lower impact on perceived causal relationships compared to bar charts and line charts.}
    On the left (a) are the simple slope regressions, and on the right (b) are slopes at each interval.
    }
    \label{fig:slope_ct}
    \vspace{-1em}
\end{figure}

Our results \add{may} support \textbf{H5}: The impact of causal priors and visualized associations on perceived causal relationships vary by chart type. See \autoref{fig:slope_ct}.
Significant interactions between \textit{chart type} and \textit{visual association} were found. Post hoc analysis was performed with estimated marginal trends comparing each \textit{chart type} pairwise, with Bonferroni corrections applied.

The Bar Chart (M=1.19, SE=0.07) and the Line Chart (M=1.36, SE=0.05) both had significantly steeper slopes compared to the Scatterplot (M=0.68, SE=0.07), ($t(14050)=5.126$, $p<.0001$, $r=.96$) and ($t(14050)=7.902$, $p<.0001$, $r=.98$) respectively.
This result indicates that when visualizing bar charts and line charts, the \textit{visualized association} \add{had} a greater influence on the \textit{perceived causal relationship} than for scatterplots.

\add{However, it is important to note that this observed impact may not be due only to the chart type.
In our study, each chart type was associated with a particular data type (time series, categorical, or continuous), such that only one chart type was shown to users for each concept pair. Therefore any differences between chart types may be due, at least in part, to differences in data types.
We therefore conclude that our results can only partly support \textbf{H5}.  \autoref{sec:discusschart} provides further discussion on this limitation.}

\subsection{Confidence}
\label{sec:analysis_conf_all}

Similar to the above analysis, we found the generalized linear regression on \textit{confidence} was statistically significant, ($F(11,14404)=24.51$, $p<.0001$, $R^2=.02$), with a significant interaction effect of \textit{causal prior} and \textit{visualized association} ($\beta=-0.27$, $p=.0002$).
This suggests a larger disagreement between \textit{causal prior} and \textit{visualized association} may lead to lower subjective \textit{confidence} and vice versa.
Please see the \href{https://osf.io/dfkv4/?view_only=f84ffbc28cdf45e5a3d68f2f1e9c8427}{\textcolor[RGB]{0,0,255}{OSF}} supplements for a more comprehensive reporting of these results.

\section{Modeling Causality Perception}
\label{sec-patterns}

In addition to the significant results reported in \autoref{sec-analysis-1} and \autoref{sec-analysis-2}, we report additional data patterns on causal priors observed in Study 1 in \autoref{sec:analysis_prior_exist}, differential data patterns observed in Study 2 in \autoref{sec:analysis_diff}, and a model for perceived causal relationship in  \autoref{sec:model_causal}.

\subsection{Causal Prior Data Patterns}
\label{sec:analysis_prior_exist}

\begin{figure}[t]
\centering
\includegraphics[width=\columnwidth]{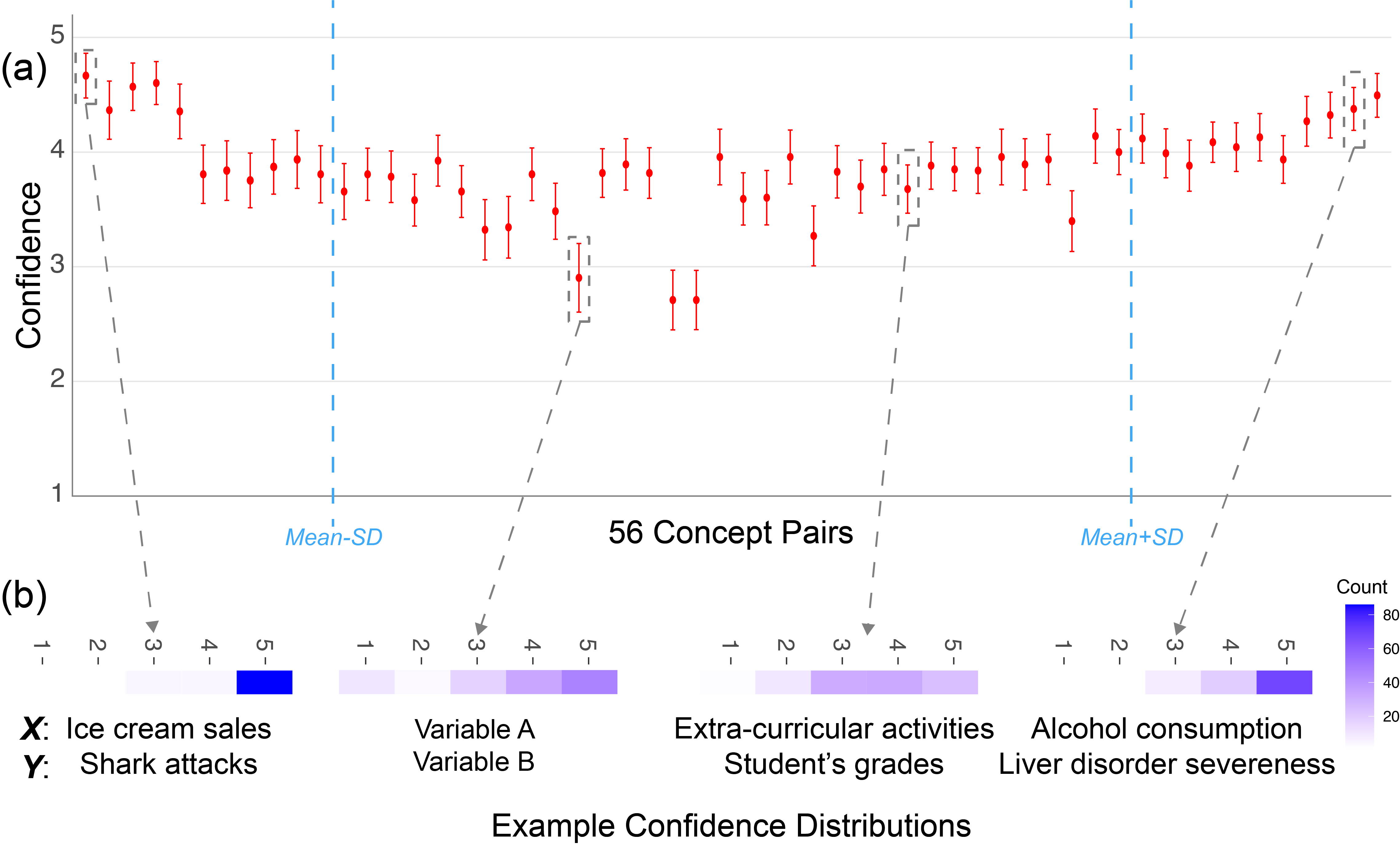}
\vspace{-1em}
\caption{ 
\add{Results of user-reported confidence from} Study 1. 
Concept pairs are ordered with increasing causal prior. The cyan lines indicate mean $\pm$ SD cyan lines, as with \autoref{fig:teaser}.
}
\label{fig:wordresultsconf}
    \vspace{-1em}
\end{figure}

\begin{figure*}[b]
\vspace{-1em}
    \centering
    \includegraphics[width=1.6\columnwidth]{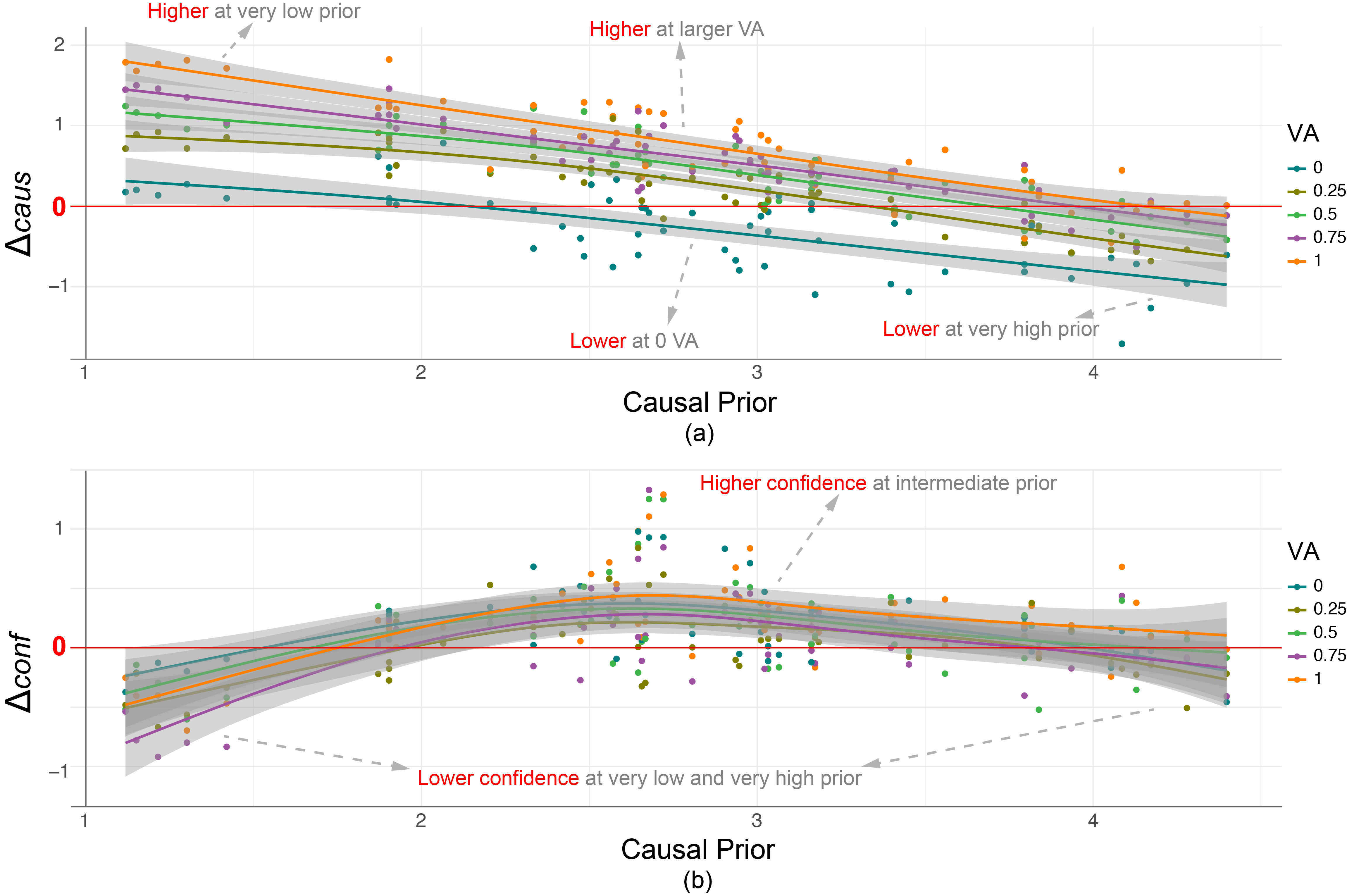}
\vspace{-0.5em}
\caption{\add{Overall results of $\Delta_{caus}$ (a) and $\Delta_{conf}$ (b), as defined by \autoref{eq:p} and \autoref{eq:dconf}.}
    The x-axis is causal prior values of ordered concept pairs following \autoref{fig:teaser}.
    Each point in this figure denotes an average value of $\Delta_{caus}$ or $\Delta_{conf}$, which is the mean difference, for each concept pair.
    The trendlines are fitted per visualized association with generalized additive models~\cite{wood2017generalized} using $\Delta_{caus}$ or $\Delta_{conf}$ calculated via all user-reported results in all concept pairs.
    The colors denote different levels of visualized association (VA) from 0 to 1, at 0.25 intervals.
    \add{(a) hints that people intend to judge higher causal relationships for concept pairs with very low causal priors, and vice versa.
    (b) shows people have lower confidence in concept pairs with both very high and low causal priors.}
    }
\vspace{-1em}
    \label{fig:diff}
\end{figure*}

\autoref{fig:teaser} (a) indicates that users' reported causal relationships for concept pairs are more tightly distributed, exhibiting smaller 95\% confidence intervals, for more extreme causal priors with low or high values.
For example, as shown in \autoref{fig:teaser} (b), when examining the causal relationship between \textit{ice cream sales} and \textit{shark attacks}, 86 users out of 92 chose a score of one, and for \textit{alcohol consumption} and \textit{liver disorder severeness}, 53 users chose a score of five and 32 a score of four.

\autoref{fig:teaser} (a) also shows that the causal relationship scores are more uniformly distributed, exhibiting larger 95\% confidence intervals across the five rating levels, for concept pairs with intermediate causal priors, i.e., within the range of mean $\pm$ SD.
For example, see the results between \textit{variable A} and \textit{variable B} in \autoref{fig:teaser} (b).

Interesting data patterns also appear in participant-reported confidence from Study 1.
\autoref{fig:wordresultsconf} (a) indicates the confidence score distributions per increasing average causal relationship order of concept pairs, and
\autoref{fig:wordresultsconf} (b) shows the confidence distribution of the same example cases.

These results indicate that for concept pairs with both very low or very high causal priors
, the confidence is distributed in the high-value range.
For example, in \autoref{fig:wordresultsconf} (b), more than 80 users scored five in confidence for \textit{ice cream sales} and \textit{shark attacks}, 64 users scored five and 18 scored four for \textit{alcohol consumption} and \textit{liver disorder severeness}.

But for concept pairs with intermediate causal priors, especially for those between mean $\pm$ SD, users' confidence becomes obviously lower.
E.g., see \textit{extra-curricular activities} and \textit{students' grades} in \autoref{fig:wordresultsconf}~(b)).

\subsection{Exploratory Analysis for Differential Data Patterns}
\label{sec:analysis_diff}

We then analyzed data patterns from Study 2 using two differential measures that calculate the difference between reported results from users who see visualizations in Study 2 with the causal priors. First, differential perceived causal relationship, $\Delta_{caus}$, is defined as follows:
\begin{equation}
\label{eq:p}
    \Delta_{caus} = PCR - CausalPrior,
\end{equation}
where $PCR$ is the perceived causal relationship from Study 2.

A second measure, differential confidence (noted as $\Delta_{conf}$), captures the corresponding difference in confidence as defined below:
\begin{equation}
\label{eq:dconf}
    \Delta_{conf} = PConf - ConfPrior
\end{equation}

In this way, we can use these two measures to assess to what extent visualizations may lead to differences in users' responses compared to those reported without seeing visualizations.
A positive $\Delta_{caus}$ value means that being exposed to the corresponding visualized data leads to an increased perceived causal relationship while a negative value refers to visualized data leading to a decreased perceived causal relationship. A positive or negative $\Delta_{conf}$ value can be interpreted similarly with respect to the difference in reported confidence.

\autoref{fig:diff} shows the overall results for $\Delta_{caus}$ and $\Delta_{conf}$ grouped by visualized association and plotted as a function of causal prior.
Given this representation, two patterns emerge that can provide further insights into our findings between causal prior and visualized associations.

First, \autoref{fig:diff} (a) suggests that visualizations may increase users' perceived causal relationships for concept pairs with low causal priors (on the leftmost side), 
while also decreasing users' perceived causal relationships on concept pairs with high causal priors (on the rightmost side).
We note, however, that this moderating effect may be related to a ``regression to the mean'' effect~\cite{barnett2005regression}.

Second, \autoref{fig:diff} (b) reveals that users' confidence may be impacted by visualized data in a way that is different from $\delta_{caus}$. 
For concept pairs with both very low and very high causal priors, users seeing visualizations tended to report lower confidence compared to users who only saw concept pairs.
However, for concept pairs with more intermediate causal prior, we found users who saw visualizations, no matter the visualized association levels, became more confident compared to those who only saw concept pairs.
\add{Overall, these results suggest that for concept pairs with middle causal priors, the impact of visualizations could be higher than those extreme cases.}

\subsection{Modeling Differential Perceived Causal Relationship}
\label{sec:model_causal}

The results from \autoref{sec:analysis_agree} and \autoref{sec:analysis_diff} indicate that both causal priors and visualized associations impact perceived causal relationships. Here we present a model designed to help characterize this relationship. 

First we calculate the difference between the visualized association (\add{denoted as $VA$ for the remainder of the paper, not to be confused with $VA$ as an abbreviation of \emph{visual analytics}}) and causal prior, $\Delta_{VA, CP}$:
\begin{equation}
\label{eq:vap}
    \Delta_{VA, CP} = VA - normalize(CausalPrior),
\end{equation}
where the normalize function maps the causal prior from [1, 5] to [0, 1] to match the range of $VA$.

\begin{figure}[t]
    \centering
    \includegraphics[width=\columnwidth]{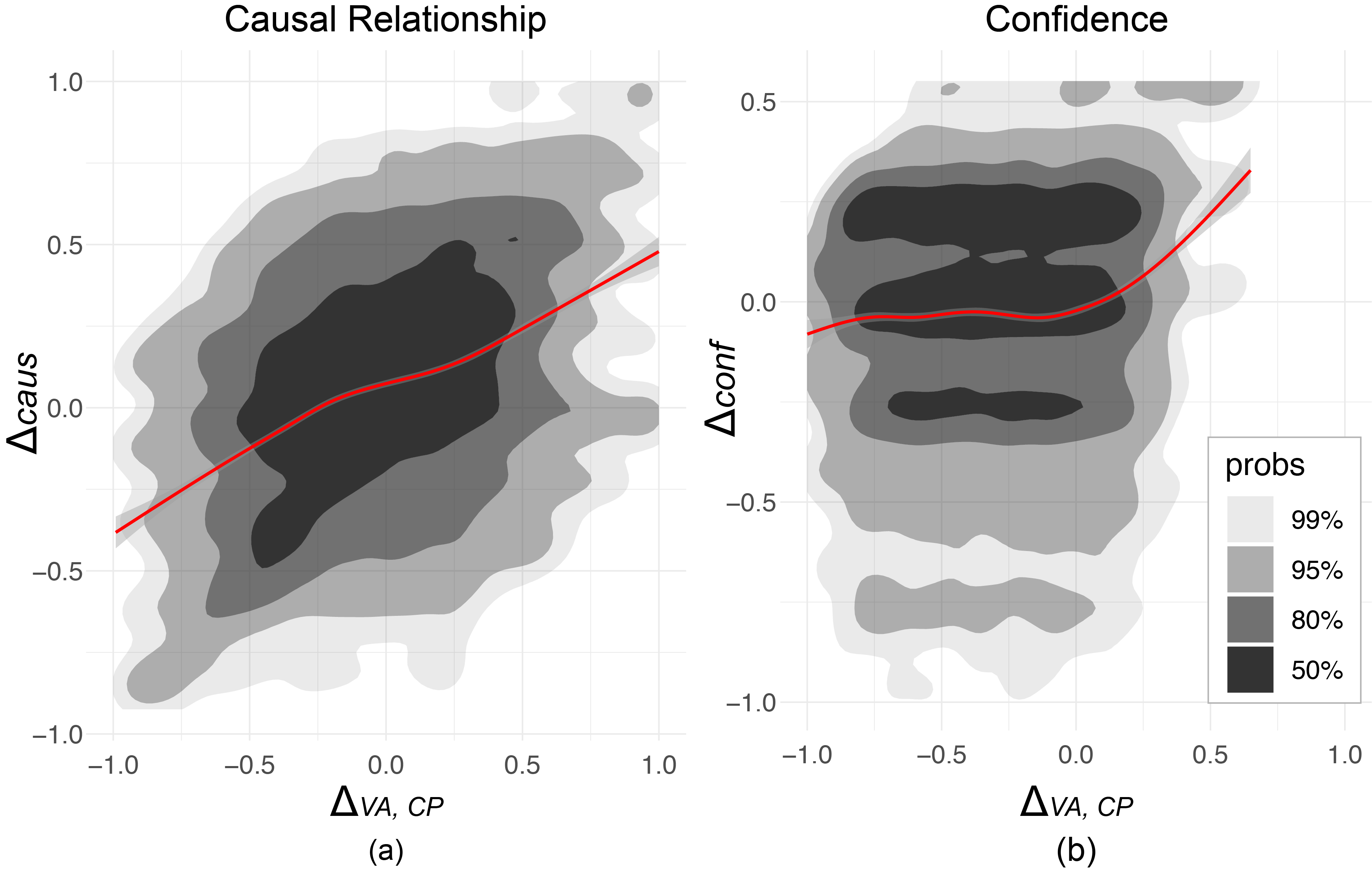}
\vspace{-1em}
    \caption{\add{Relationships between $\Delta_{VA, CP}$ and $\Delta_{caus}$ (a) and $\Delta_{conf}$ (b).
    The $x$-values denote the values of $\Delta_{VA, CP}$, while $y$-values show how much $\Delta_{caus}$ and $\Delta_{conf}$ change with respect to $x$.}
    The heatmaps show the densities of all data points in each graph; both charts use the gray color scale shown on the bottom-right, indicating the proportion of the number of points shown within each color.
    The trendlines were fit with generalized additive models~\cite{wood2017generalized} on 95\% confidence intervals.}
    \label{fig:model}
    \vspace{-1em}
\end{figure}

\autoref{fig:model} shows comparisons of the distributions of $\Delta_{VA, CP}$ to $\Delta_{caus}$ in (a) and $\Delta_{conf}$ in (b).
The ranges of $\Delta_{caus}$ and $\Delta_{conf}$ were normalized to [0, 1] for easier comparison in this figure.

In \autoref{fig:model} (a), the densities clustered around the coordinate origin indicate that when $\Delta_{VA, CP}$ is near 0, $\Delta_{caus}$ also tends to be near 0.
Furthermore, the trendline shows $\Delta_{VA, CP}$ exhibits a close to linear relationship with $Delta_{caus}$
\add{, suggesting a positive disagreement would lead to users' higher judgment of causality and vice versa,} which aligns with our findings.
\add{Besides, we fit a linear regression model to the distribution between $\Delta_{VA, CP}$ and $Delta_{caus}$, with the result:}
\begin{equation}
\label{eq:causalmodel}
    \Delta_{caus} \sim 0.37*\Delta_{VA, CP},
\end{equation}
with an intercept at 0.06, $p<.0001, R^2=.22$.
\add{This equation indicates that}, on average, the difference between visualized association and causal prior will contribute \add{0.37} of the difference in the final perceived causal relationship.
\add{Further, beyond the overall trend that can be captured by this model, we noticed that there exist more complex data patterns.
For example, we found some data points are clustered at around \emph{(-1.0, -1.0)} and \emph{(1.0, 1.0)} indicating an extreme disagreement may have a more severe impact than the general data pattern.
We also found a densely clustered pattern near \emph{x=0}, suggesting many users still keep their ratings near the original causal priors.}

In \autoref{fig:model} (b), however, the densities cluster at different levels of $\Delta_{conf}$, indicating that we cannot simply estimate how much users' confidences may change using the differences between visualized association and causal prior.
This aligns with the differential patterns of confidence discussed in \autoref{sec:analysis_diff} and \autoref{fig:diff} (b).
\add{Meanwhile, it's obvious that the densely clustered data points are mostly above zero and near horizontal lines, possibly indicating visualizations are more likely to increase users' confidence and the visualized associations may have a lower impact.}
\add{These initial patterns indicate a need to better justify them in future research.}

\section{Discussion}
\label{sec-discussion}
Our study investigated the impact of underlying causal priors and visualized associations on causal strength ratings between concept pairs.
We find that human cognition of causal inference is impacted by both preexisting causal priors and the displayed visualizations.

\subsection{Reflections on Previous Findings}

Our results indicate many insights and relations related to underlying causal priors between concept pairs, associations conveyed via visualization, chart types, and human cognition of causal inference.
The findings complement observations from past work focusing on related elements of visualization usage and visual causal inference.

\subsubsection{Underlying Causal Prior in Mind}
The first major finding indicates that people have preexisting causal priors (see \autoref{sec-analysis-1}).
This insight aligns with the results from previous studies, such as Ferstl~et al.\cite{ferstl2011implicit} and Goikoetxea et al.~\cite{goikoetxea2008normative}, which found that there exist implicit causality norms from particular corpora of English and Spanish verbs respectively.
Our results differ in that we gain insights between concepts represented by noun phrases, and because our public corpus may be more applicable for certain analytical tasks as most of the tested concepts are gathered from meaningful variable pairs from widely applied datasets.

\subsubsection{Association in Visualizations}
Furthermore, our results demonstrate that the correlations broadly represented in visualizations significantly impact human causal inference.
The analysis provided further evidence built upon existing insights on how correlations in visualizations can impact the human perception of causal relations~\cite{cleveland1988shape, harrison2014ranking, kay2015beyond, rensink2010perception, rensink2017nature}.
However, due to the significant impact of causal priors, we found that causal inference can be more complicated than just correlation estimation, thus the perceived causal relationship cannot be simply modeled with either a linear or log-linear model of visualized correlations such as those employed by many correlation estimation studies~\cite{rensink2010perception, harrison2014ranking, kay2015beyond}.

\subsubsection{Estimation of Causal Relationships}

We found that causal priors have a significant impact on the cognition of causal inference from visualizations.
Although there exists limited previous research focus on priors in visual causal inference, our findings align with existing studies that focus on underlying relevant and background knowledge combined with correlation.
For example, the impact of underlying causal priors shows a similar tendency to many existing studies which found that underlying background knowledge significantly impacts people's understandings of visualizations~\cite{larkin1987diagram, liu2010mental, padilla2018decision}.
Additionally, this impact also aligns with Xiong~et al.'s findings that people's beliefs impact their estimation of correlations~\cite{xiong2022seeing}.

Our results further support an important insight from Kale et al.~\cite{kale2021causal}.
Their study found that users' perception of causal inference tends to overestimate or underestimate the ground truth causal relations built upon the causal support model.
As shown in \autoref{sec:analysis_diff}, our findings align with their insight and provide evidence that for concept pairs, such cases more often appear with either very low or very high causal priors. 
This insight may also align with psychology phenomena such as the regression to the mean effect~\cite{barnett2005regression}, which states that human predictions on extreme values, whether high or low, tend to be near the intermediate.
However, our study is not designed to test psychological effects, further experiments will need to be conducted to validate these assumptions.
For intermediate causal prior concept pairs, users' causality perception is largely impacted by visualizations.
Additionally, our findings also show evidence that users' confidence varies based on concept pairs with different causal priors and visualized associations.

\subsubsection{Chart Type}
\label{sec:discusschart}
Finally, our results suggest that chart type \add{may} impact causal inference.
We found that the perceived causal association for bar charts and line charts had a greater difference from the causal prior than for scatterplots, which aligns with prior studies~\cite{xiong2019illusion, wang2024empirical}.
Xiong et al.~\cite{xiong2019illusion} explained these differences could be due to the different aggregation levels of these charts.
\add{However, unlike their study, which tested different chart types for the same datasets, each chart type in our study was only used for a particular data type (e.g., scatterplots for continuous data).
As a result, only one chart type was used for each concept pair. This limits our ability to disentangle the effects of chart type from the potential effects of data type.
Future studies should assess how data type, chart type, and other potential factors %
}

\subsection{Implications for Design}
\label{sec:design}

In the light of existing perception experiments~\cite{xiong2019illusion, kale2021causal, kaul2021improving, wang2024empirical}, our study confirms many insights and goes further in interpreting the cognition process of causal inference by connecting underlying causal priors and visualized associations.
The results further lead to heuristic implications for the guidelines of future visual causal inference research and system design, resulting in the following key points.

\paragraph{Causal priors:}
Designers should recognize that users have pre-existing causal priors for many causal relationships.
For specific analytical usage scenarios that may share a large number of similar variables, designers can collect analysts' underlying causal priors for those concept pairs in advance to better guide their interpretation of visualizations.
This understanding could lead to the development of adaptive visualization systems that tailor the presentation of data based on the user’s established causal prior knowledge base.
\add{However, there remain challenges to properly and precisely accessing the causal priors of target users, which may be hard and unrealistic to assume.
Therefore,}
future design and research should also consider the causal priors of the users when presenting causal information, and provide ways to elicit, validate, and update them.

\paragraph{Disagreement impact:}
Disagreement between a user’s causal prior and the visualized association can lead to cognitive dissonance on overestimation or underestimation of causal relationships.
Since the visualized association can be pre-calculated by visual analytics systems, we would recommend adding functionalities of disagreement checks using our suggested model in \autoref{sec:model_causal}, if the causal prior has been collected in advance.
\add{It's also important for future research to find a balance for such impact to avoid undervaluing human knowledge and intuition and overvaluing the visualized data.}

\paragraph{Visualization types:} 
Previous work has shown that choices of chart type play a pivotal role in the interpretation of causal relationships, and our results seem to agree.
Developers should be aware the overall cognitive process of causal inference that chart types may evoke should be an important consideration of effective visualization design.
Future design efforts should focus on identifying which types of visual representations are most conducive to accurate causal inference.

\subsection{Limitations and Future Directions}
\label{sec:limit}

\paragraph{Data corpus:} 
Although we collected concept names from variables from widely applied datasets, to achieve an efficient MTurk study design our corpus only contains 56 concept pairs.
This introduces some limitations on the distribution of causal priors. For example, we didn't find any concept pair that has an extremely high average causal prior, e.g., larger than 4.6, whereas the highest causal prior we found was around 4.5.
Future research should focus on enlarging the concept pair corpus to better cover all possible distributions of underlying causal priors.
In addition, we tested one direction in the causal questions, by predetermining causal factors and outcome concepts (\autoref{sec:stimuli}).
Further experiments could consider the impact of different directions between concept names in causal questions.

\paragraph{Visualization stimuli design:} 
Additionally, to simplify the study design, we only considered three typical chart types, following previous work~\cite{quadri2024do, wang2024empirical, xiong2019illusion}, and did not test more complex visual encodings, such as aggregation levels within each chart type.
Therefore, although we confirmed many existing insights regarding these charts, we cannot directly address insights related to other visual encodings, such as how text tables were found to perform similarly to bar charts~\cite{kale2021causal}, and increasing aggregation levels in specific chart types were found to improve the ability to convey causality~\cite{xiong2019illusion}.
Moreover, our chart types were associated with data types, so we are unable to decouple the potential impacts of data types and chart types, as discussed in \autoref{sec:discusschart}.
We also did not test visualizations showing negative associations, which could have a different impact on human perception in certain cases~\cite{beheshitha2016role}.
\add{In addition, we removed chart axes since determining appropriate axis value complicated the study design. However, showing such values may also impact users' perceptions.}
We plan to incorporate more design factors in visualizations such as additional chart types, varying visual encodings, and negative associations in future work.

\paragraph{Expertise and visualization literacy:} 
Our study did not aim to assess the impact of visualization expertise,
and was conducted on the general population of MTurk users.
Our results can therefore not provide insights related to existing perception experiments specifically designed for novices~\cite{grammel2010information, burns2023we, peck2019data, burns2020evaluate} or experienced users~\cite{battle2019characterizing, gotz2009behavior, gotz2008characterizing, stolper2014progressive}.
Future work could focus on examining how our findings may be affected by novice or expert populations.

\section{Conclusion}
\label{sec-conclusion}

This paper aims to provide insights on the human cognition process of causal inference from visualizations, and provides new perspectives on guidelines from graphical perception
for visual causal inference.
Our study has illuminated the significant role of causal priors, i.e., the preconceived notions about causality between concept pairs, in shaping users’ interpretations of visualized data.
The reported empirical evidence underscores the influence of causal priors, which, when combined with evidence of associations in visualized data, can significantly impact cognitive processes related to causal inference.
\add{The results additionally suggest that chart type may also have an impact.}
The introduction of a model capturing the differential patterns in perceived causal relationships caused by causal priors and visualized associations offers a novel perspective on how these elements converge to affect users’ causality judgments.
Furthermore, our research contributes to the field by providing a comprehensive dataset of causal priors associated with concept pairs and visualizations.
This dataset not only serves as a benchmark for future studies but also aids in the development of heuristic-based design guidelines aimed at enhancing visual design choices to better support visual causal inference based on specific variables in data.
In light of our findings, we advocate for heightened awareness among visualization designers regarding the potential for causal priors to lead to the impact of perceived causal relationships in the visualized data.
By integrating these heuristic-based guidelines into visualization design, we can facilitate more accurate interpretations of visualized data, thereby improving decision-making processes across various domains.

\acknowledgments{
We thank the reviewers for their insightful comments.
This material is based upon work supported by the National Science Foundation under Grant No. 2211845.
}

\bibliographystyle{abbrv-doi-hyperref}

\bibliography{main}

\end{document}